\documentstyle [12pt,epsfig]{article} 
\makeatletter
\@addtoreset{equation}{section}
\makeatother

\newcommand{\ba}{\begin{eqnarray}}
\newcommand{\ea}{\end{eqnarray}}

\begin{document}
\newcommand{\BS}{\bigskip}
\newcommand{\SECTION}[1]{\BS{\large\section{\bf #1}}}
\newcommand{\SUBSECTION}[1]{\BS{\large\subsection{\bf #1}}}
\newcommand{\SUBSUBSECTION}[1]{\BS{\large\subsubsection{\bf #1}}}
\newcommand {\rb}  {\overline{r}}
\newcommand {\sbx}  {\overline{s}}
\newcommand {\vb}  {\overline{v}}
\newcommand {\ab}  {\overline{a}}
\newcommand {\sbf}  {\overline{s}_f}
\newcommand {\rbf}  {\overline{r}_f}
\newcommand {\vbf}  {\overline{v}_f}
\newcommand {\abf}  {\overline{a}_f}
\newcommand {\sbq}  {\overline{s}_q}
\newcommand {\vbq}  {\overline{v}_q}
\newcommand {\abq}  {\overline{a}_q}
\newcommand {\rbQ}  {\overline{r}_Q}
\newcommand {\vbQ}  {\overline{v}_Q}
\newcommand {\abQ}  {\overline{a}_Q}
\newcommand {\rbl}  {\overline{r}_l}

\newcommand {\vbl}  {\overline{v}_l}
\newcommand {\abl}  {\overline{a}_l}
\newcommand {\sbl}  {\overline{s}_l}

\newcommand {\rbc}  {\overline{r}_{\rm{c}}}
\newcommand {\vbc}  {\overline{v}_{\rm{c}}}
\newcommand {\abc}  {\overline{a}_{\rm{c}}}
\newcommand {\sbc}  {\overline{s}_{\rm{c}}}
\newcommand {\rbb}  {\overline{r}_{\rm{b}}}
\newcommand {\vbb}  {\overline{v}_{\rm{b}}}
\newcommand {\abb}  {\overline{a}_{\rm{b}}}
\newcommand {\sbb}  {\overline{s}_{\rm{b}}}
\newcommand {\gbl}  {\overline{g}_{\rm{b}}^L}
\newcommand {\gbr}  {\overline{g}_{\rm{b}}^R}
\newcommand {\gcl}  {\overline{g}_{\rm{c}}^L}
\newcommand {\gcr}  {\overline{g}_{\rm{c}}^R}
\newcommand {\Afbf}  {A_{FB}^{0,\rm{f}}}
\newcommand {\Afbl}  {A_{FB}^{0,\ell}}
\newcommand {\Afbc}  {A_{FB}^{0,\rm{c}}}
\newcommand {\Afbb}  {A_{FB}^{0,\rm{b}}}
\newcommand {\tpol}  {$\tau$-polarisation}
\newcommand {\alps}  {\alpha_s(M_\rm{Z})}
\begin{titlepage}
%
{\bf Revised Version 14/10/99}
\hspace*{2cm} {UGVA-DPNC 1999/7-183 July 1999}
\newline
\hspace*{10cm} {hep-ex/9907018}
\begin{center}
\vspace*{2cm}
{\large \bf
A Study of the LEP and SLD Measurements of $A_{\rm{b}}$}

\vspace*{1.5cm}
\end{center}
\begin{center}
{\bf J.H.Field and D.Sciarrino}
\end{center}
\begin{center}
{ 
D\'{e}partement de Physique Nucl\'{e}aire et Corpusculaire
 Universit\'{e} de Gen\`{e}ve . 24, quai Ernest-Ansermet
 CH-1211 Gen\`{e}ve 4.
}
\end{center}
\vspace*{2cm}
\begin{abstract}
A systematic study is made of the data dependence of the parameter $A_{\rm{b}}$,
that, since 1995, has shown a deviation from the Standard Model prediction
of between 2.4 and 3.1 standard deviations. Issues addressed include:
the effect of particular measurements, values found by individual experiments,
LEP/SLD comparison, and the treatment of systematic errors. The effect,
currently at the 2.4$\sigma$ level, is found to vary in the range
from 1.7$\sigma$ to 2.9$\sigma$ by excluding marginal or particularly
sensitive data. Since essentially the full LEP and SLD Z decay data sets
are now analysed the meaning of the deviation, (new physics, or marginal
statistical fluctuation) is unlikely to be given by the present
generation of colliders.            
\end{abstract}
\vspace*{1cm}

\end{titlepage}
\SECTION{\bf{Introduction}}
As has been recently pointed out in the literature~[1-4],
 the analysis
of the precision data on the decays $\rm{Z} \rightarrow \rm{f} \overline{\rm{f}}$ from
LEP and SLD has shown good agreement with the predictions of the
Standard Electroweak Model (SM)~\cite{x5} with the exception of the
 parameter $A_{\rm{b}}$
defined as:
\begin{equation}
    A_{\rm{b}} = \frac{2 (\sqrt{1-4 \mu_{\rm{b}}}) \rbb}{1-4 \mu_{\rm{b}}+(1+2
 \mu_{\rm{b}}) \rbb^2}
\end{equation}
where 
\[ \rbb =\vbb/\abb \]
Here $\vbb$ and $\abb$ are the effective b quark coupling constants and
 $\mu_{\rm{b}} = (\overline{m}_{\rm{b}}(M_{\rm{Z}})/M_{\rm{Z}})^2 \simeq 1.0 \times
 10^{-3}$~\cite{x6}.
 Since 1995, the LEP+SLD average value of $A_{\rm{b}}$ has differed from the
 SM prediction of 0.935~\cite{x1} by between 2.4 and 3.1 standard 
 deviations. The evolution with time of the LEP+SLD average value
 of $A_{\rm{b}}$ is shown in Table 1 and Fig.1 [1,7-13]. 
 It is important to note that, in the SM, the prediction
 for $A_{\rm{b}}$ is essentially a fixed number with no significant dependence
 on the values of the masses of of the top quark or the Higgs boson (see
 Figs 5-7 below).  
 Combining the $A_{\rm{b}}$ measurement with that of $R_{\rm{b}}$, which 
 shows relatively good agreement with the SM, enables the effective b quark
 couplings $\vbb$, $\abb$ or $\gbl$, $\gbr$ to be extracted~[2-4].
 When this is done, the largest deviation from the SM prediction is
 found to be in the right handed effective coupling $\gbr$ which is about
 40$\%$ and three standard deviations higher than the SM prediction.
 \par The aim of the present note is a thorough study of the data 
  dependence of the LEP+SLD average value of $A_{\rm{b}}$. Important questions
  concern the consistency of individual measurements, and the effect
  of one or a few `outlying' measurements on the average. At SLD the
  parameter $A_{\rm{b}}$ is measured directly from the forward/backward, left/right
  asymmetry of tagged b quarks. Three different types of measurement are
  made. The b quarks are tagged using a decay vertex and the jet charge,
  a semi-leptonic weak decay, or a $\rm{K}^{\pm}$ tag~\cite{x12}. The LEP value
  of $A_{\rm{b}}$ is instead derived from the $Z$-pole forward/backward charge
  asymmetry, related to $A_{\rm{b}}$ by the expression:
\begin{equation}
\Afbb  =  \frac{3}{4} A_{\rm{e}} A_{\rm{b}}
\end{equation}
 where $A_e$ is the parameter defined similarly to $A_{\rm{b}}$ (Eqn.(1.1)) for
the electron. In general lepton universality
i.e. $A_{\ell} = A_{\rm{e}} = A_{\mu} = A_{\tau}$, is assumed. Each of the four LEP
 experiments measures $A_{FB}^{0,\rm{b}}$ using either a lepton tag or the
 combination of decay vertex and jet charge measurements. Thus there
 are eight separate (though not completely uncorrelated) LEP measurements
 of $\Afbb$. Using the 
 LEP+SLD average value of $A_{\ell}$ ($A_{\ell} = 0.1490 \pm 0.0017$) and Eqn.(1.2)
the corresponding values of $A_{\rm{b}}$ for each LEP experiment and each 
 analysis method may be calculated. These results are shown, together
 with the three direct SLD measurements, in Table 2 and Fig.2.
  The data shown are the most recent (Spring 1999) available at the
  time of writing. They are essentially the same as those presented
  at the 1998 Vancouver conference~\cite{x12}  except for the recent important
  update~\cite{x13} of the SLD jet charge measurement
 which yields an SLD average value of $A_{\rm{b}}$ that is 
  consistent, at the one standard deviation level, with the SM
 prediction.

\par Because the LEP value of $A_{\rm{b}}$ depends directly on the LEP+SLD
 average value of $A_{\ell}$, it is of interest to compare the different 
 measurements of this quantity. Each of the four LEP experiments 
 measures $A_{\ell}$ either via the forward/backward leptonic charge
 asymmetry:    
\begin{equation}
 A_{\ell}  = \sqrt{\frac{4A_{FB}^{0,\ell}}{3}}~,~~~(l=\ell,\mu,\tau)
\end{equation}
or by the analysis of \tpol. The angular average of the \tpol~ 
measures $A_{\tau}$, whereas the angular distribution of the
 polarisation is also sensitive to $A_e$. Combining, for each LEP 
 experiment, under the assumption of lepton universality,
 the measurements of $A_{\tau}$ and $A_e$, and including $A_e$
 as measured at SLD by the left/right electron beam polarisation 
 asymmetry, leads to the nine independent measurements of $A_{\ell}$ shown
 in Table 3 and Fig.3.  
 \par Very good consistency can be seen in Table 2 and Fig.2 between
  the 11 different measurements of $A_{\rm{b}}$ 
  ($\chi^2/dof = 4.5/10, CL=92 \%$ for
  consistency of the measurements with their weighted mean). The LEP
  and SLD average values agree within 0.2$\sigma$. As noted also for
 the 1996 
  data set~\cite{x4}, the mutual consistency of the different $A_{\ell}$
  measurements is somewhat less satisfactory.
  Although the $\chi^2$ test gives: $\chi^2/dof = 10.7/8, CL=22 \%$
  which is acceptable, three measurements
  (OPAL $A^{0,\ell}_{FB}$ and the \tpol~ measurements of DELPHI and OPAL)
  all show negative deviations of 1.5$\sigma$ or more
 from the weighted average value. In contrast,
   all the positive deviations are $\le$ 1$\sigma$. The average value
   of $A_{\ell}$, and hence the derived LEP value of $A_{\rm{b}}$ is thus
    sensitive to the inclusion or exclusion of these data, as will be
  discussed below. The situation concerning the consistency of the
  \tpol~measurements, both with each other, and with the other
 determinations of $A_{\ell}$, discussed in detail for the 1996
 data set in reference~\cite{x4}, has recently been improved by the new, more
 precise,\mbox{ ALEPH measurement (see Fig 3).}
\begin{table}
\begin{center}
\begin{tabular}{|c|c|c|c|} \hline
 Year & Reference & $A_{\rm{b}}$ & Deviation ($\sigma$) from SM \\ \hline
 1993 & [7] & 0.925(56) & -0.18 \\
 1994 & [8] & 0.934(48) & -0.02 \\
 1995 & [9] & 0.871(27) & -2.4 \\
 1996 & [1] & 0.867(22) & -3.1 \\
 1997 & [10] & 0.877(23) & -2.5 \\
 1998 & [11] & 0.878(19) & -3.0 \\
 1999 & [12,13] & 0.894(17) & -2.4 \\ \hline
\end{tabular}
\caption[]{ The time evolution of the
 LEP+SLD average values of $A_{\rm{b}}$.}
\end{center}
\end{table}
\begin{table}
\begin{center}
\begin{tabular}{|c|c|} \hline
\multicolumn{2}{|c|}{ SLD }  \\ \hline
 Jet Ch  & 0.882 $\pm$ 0.020 $\pm$ 0.029 (0.035) \\
 Lepton  & 0.924 $\pm$ 0.032 $\pm$ 0.026 (0.041) \\
 $\rm{K}^{\pm}$ tag  & 0.855 $\pm$ 0.088 $\pm$ 0.102 (0.134) \\ \hline
\multicolumn{2}{|c|}{LEP}  \\ \hline
 A Lepton  & 0.908 $\pm$ 0.041 $\pm$ 0.020 (0.046) \\
 D Lepton  & 0.904 $\pm$ 0.057 $\pm$ 0.026 (0.063) \\
 L Lepton  & 0.869 $\pm$ 0.055 $\pm$ 0.030 (0.063) \\
 O Lepton  & 0.851 $\pm$ 0.038 $\pm$ 0.021 (0.043) \\ \hline
 A Jet Ch  & 0.953 $\pm$ 0.037 $\pm$ 0.029 (0.047) \\
 D Jet Ch  & 0.898 $\pm$ 0.042 $\pm$ 0.021 (0.047) \\
 L Jet Ch  & 0.806 $\pm$ 0.106 $\pm$ 0.051 (0.118) \\
 O Jet Ch  & 0.898 $\pm$ 0.047 $\pm$ 0.037 (0.060) \\ \hline
 WA SLD & 0.908(27) \\
 WA LEP & 0.885(22) \\
 WA LEP+SLD  & 0.894(17) \\ \hline
\end{tabular}
\caption[]{ The different LEP and SLD measurements of $A_{\rm{b}}$.
 The first error quoted is statistical, the second systematic.
 The quadratic sum of these errors is given in parentheses. `WA'
 denotes Weighted Average. }
\end{center}
\end{table}
\begin{figure}[htbp]
\begin{center}\hspace*{-0.5cm}\mbox{
\epsfysize10.0cm\epsffile{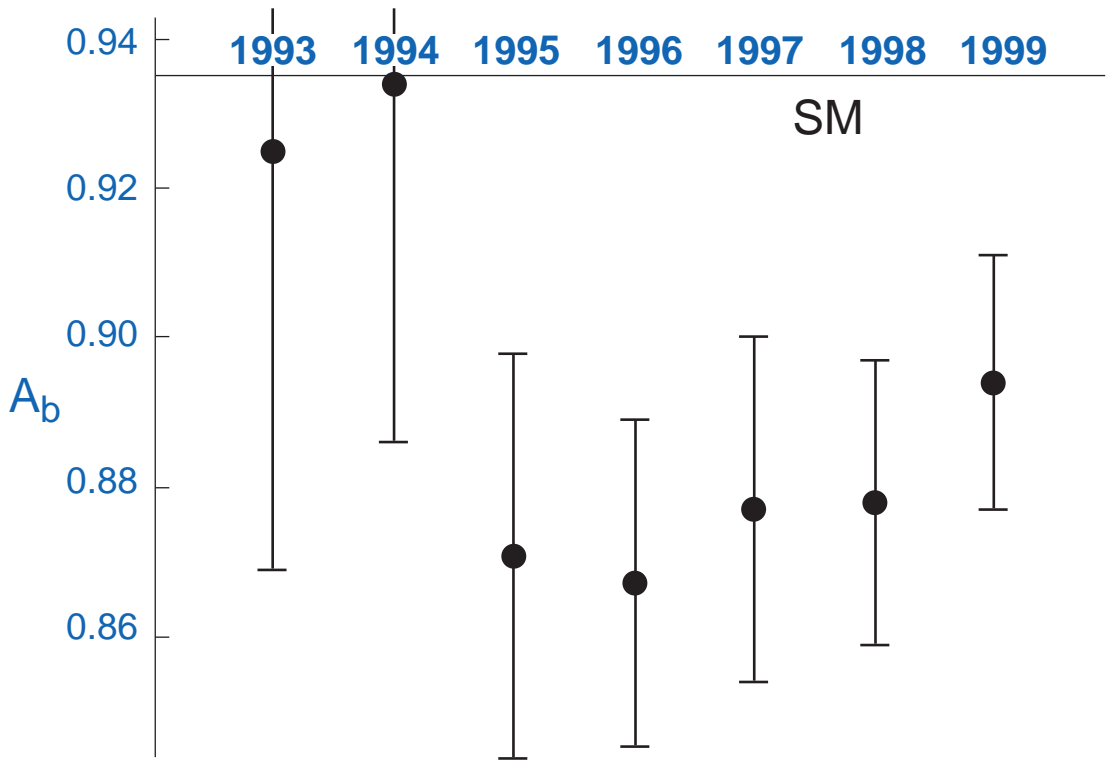}}
\caption{ The time evolution of the LEP+SLD
average value of $A_{\rm{b}}$. The
horizontal line shows the Standard Model prediction $A_{\rm{b}} = 0.935$.}
\label{fig-fig1}
\end{center}
\end{figure}
\begin{figure}[htbp]
\begin{center}\hspace*{-0.5cm}\mbox{
\epsfysize10.0cm\epsffile{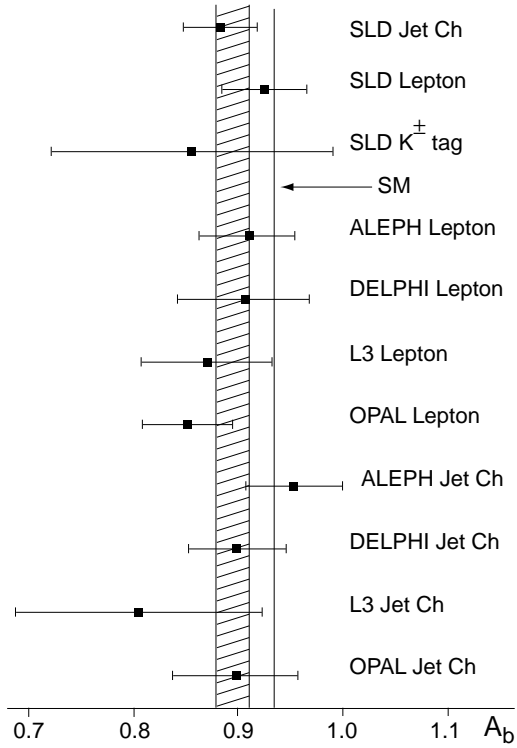}}
\caption{ LEP and SLD measurements of $A_{\rm{b}}$. The
vertical line shows the Standard Model prediction $A_{\rm{b}} = 0.935$. The
 hatched vertical band shows the weighted average value
 $\pm 1\sigma$. }
\label{fig-fig2}
\end{center}
\end{figure}
\begin{table}
\begin{center}
\begin{tabular}{|c|c|} \hline
\multicolumn{2}{|c|}{LEP $A_{FB}^{0,\ell}$}  \\ \hline
 ALEPH  & 0.1501(70) \\
 DELPHI & 0.1579(78) \\
 L3     & 0.1579(106) \\
 OPAL   & 0.1371(80) \\ \hline
\multicolumn{2}{|c|}{LEP \tpol}  \\
\hline
 ALEPH & 0.1475(46) \\
 DELPHI & 0.1369(79) \\
 L3     & 0.1558(83) \\
 OPAL   & 0.1318(100) \\
\hline
\multicolumn{2}{|c|}{SLD  $A_{LR}$}  \\
\hline
\multicolumn{2}{|c|}{ 0.1504(23)}  \\
\hline
\multicolumn{2}{|c|}{ WA LEP+SLD}  \\
\hline
\multicolumn{2}{|c|}{ 0.1490(17)} \\
\hline
\end{tabular}
\caption[]{ The different LEP and SLD measurements of $A_{\ell}$.
 The errors (quoted in parentheses) are the quadratic sums of 
 statistical and systematic errors.}
\end{center}
\end{table}
\begin{figure}[htbp]
\begin{center}\hspace*{-0.5cm}\mbox{
\epsfysize10.0cm\epsffile{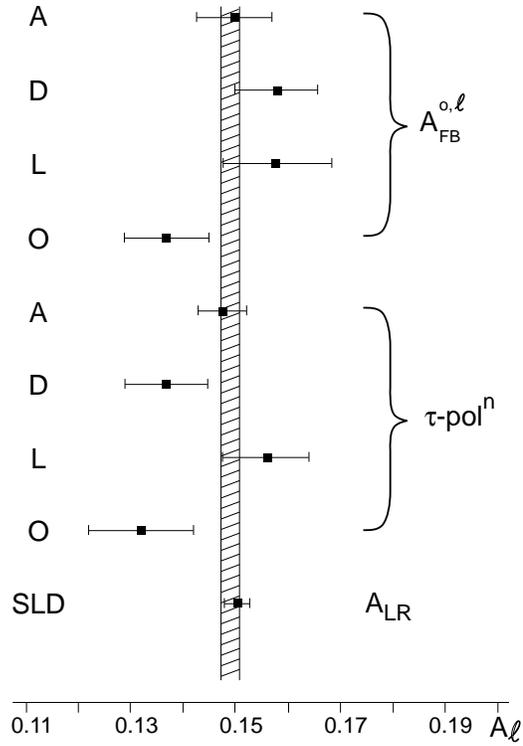}}
\caption{ LEP and SLD measurements of $A_{\ell}$.
 The hatched vertical band shows the weighted average value
 $\pm 1\sigma$. }
\label{fig-fig3}
\end{center}
\end{figure}
\begin{figure}[htbp]
\begin{center}\hspace*{-0.5cm}\mbox{
\epsfysize10.0cm\epsffile{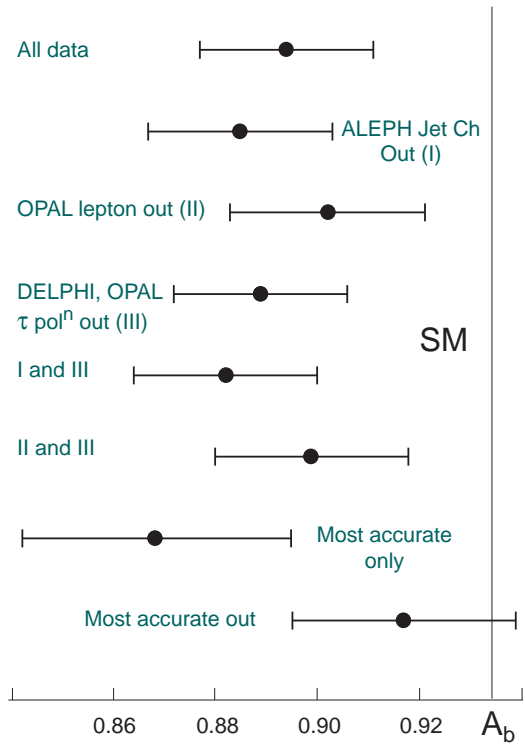}}
\caption{ Data sensitivity of the $A_{\rm{b}}$ average. The
vertical line shows the Standard Model prediction $A_{\rm{b}} = 0.935$.}
\label{fig-fig4}
\end{center}
\end{figure} 
\SECTION{\bf{Effect of Individual Measurements on the Average
 Value of $A_{\rm{b}}$}}
\begin{table}
\begin{center}
\begin{tabular}{|c||c|c|c|c|} \hline
Condition & $A_{\rm{b}}$ & Dev($\sigma$) WA & Dev($\sigma$) SM & CL SM($\%$) \\
\hline \hline
All data & 0.894(17) & 0.0 & -2.4 & 1.6 \\ \hline
ALEPH Jet Ch  &  &  &  &   \\
 out  (I)  & 0.885(18) & -0.5 & -2.8 & 0.51 \\ \hline
OPAL lepton  &  &  &  &   \\
 out  (II)  & 0.902(19) & 0.42 & -1.7 & 9.0 \\ \hline
DELPHI, OPAL &  &  &  &   \\
$\tau$ pol$^n$ out &  &  &  &   \\
  (III)  & 0.890(17) & -0.24 & -2.6 & 0.93 \\ \hline
 I and III  & 0.882(18) & -0.67 & -2.9 & 0.37 \\ \hline
 II and III  & 0.899(19) & 0.26 & -1.9 & 5.7 \\ \hline
Most accurate  &  &  &  &   \\
measurements only  & 0.868(27) & -0.96 & -2.5 & 1.2 \\ \hline
Exclude most  &  &  &  &   \\
 accurate measurements  & 0.917(22) & 1.35 & -0.82 & 41.3 \\ \hline
\end{tabular}
\caption[]{ Data sensitivity of the $A_{\rm{b}}$ average.
 Deviations from the Weighted Average (WA) and the Standard Model
 (SM) are shown, as well as the Confidence Level (CL) for agreement
 with the SM.}
\end{center}
\end{table}
 In this Section the sensitivity of the $A_{\rm{b}}$ value to 
 the different data contributing to the world average is
  examined. The results of this study are presented in Table 4.
  The ALEPH jet charge $A_{\rm{b}}$ value is the only one that lies 
  above the SM prediction. The probability that ten or more 
  out of eleven measurements of a quantity all lie either
   above or below the expected value is 1.2$\%$. Removing the ALEPH
 jet charge measurement increases the deviation from
 -2.4$\sigma$ to -2.8$\sigma$.
 The $\Afbb$ measurement with the largest weight in
 reducing the average value of $A_{\rm{b}}$ is the OPAL lepton measurement.
 Excluding this datum gives $A_{\rm{b}}=0.902(19)$ only 1.7$\sigma$ below
 the SM prediction. This single measurement gives, therefore, a 
 significant contribution to the overall deviation of $A_{\rm{b}}$. As
 discussed in detail in Ref.~\cite{x4}, apparent inconsistencies
 exist between the ~\tpol~ measurements of $A_{\ell}$ by the different
 LEP experiments. Currently two measurements (ALEPH and L3) show 
 good agreement with the Weighted Average (WA) value, whereas the
 other two (OPAL and DELPHI) show rather large (1.5-2.0$\sigma$)
 deviations
 as shown in Fig.3 and Table 3. Removing the latter measurements 
 gives a small increase of the deviation from the SM
 to -2.6$\sigma$. 
 Removing both the ALEPH and the DELPHI~\tpol~ measurements and
 the ALEPH jet charge $A^{0,b}_{FB}$ result
 increases the deviation to -2.9$\sigma$, whereas removing the
 same ~\tpol~ measurements and the OPAL lepton $\Afbb$
 result reduces the deviation to -1.9$\sigma$. Thus exclusion
 of `marginal' data results in a variation of the $A_{\rm{b}}$ deviation
 from -1.7$\sigma$ to -2.9$\sigma$ as compared to the all data 
 deviation of -2.4$\sigma$. One may remark however that, in general,
 removal of the data with the largest deviations from the average
 values (OPAL for $A^{0,l}_{FB}$, DELPHI and OPAL ~\tpol~ for $A_{\ell}$;
 ALEPH jet charge for $\Afbb$) tends to increase, not 
 decrease the deviation from the SM. As mentioned above, the single
 measurement with the largest weight in the deviation is the OPAL
 lepton measurement of $\Afbb$.
\par The average $A_{\rm{b}}$ value given by the LEP jet charge measurements,
 0.913(28), shows good agreement with the SM prediction and is
 somewhat higher than the similar average of the lepton measurements,
 0.880(26). However, the difference is mainly due to the high value 
 of ALEPH measurement. Excluding this gives, for the jet charge 
 average, 0.890(35), which agrees with the lepton average within
 0.2$\sigma$.
\par In the last two rows of Table 3 are shown the results of 
calculating $A_{\rm{b}}$ using either (i) only the measurements of each
 raw observable with the smallest total error, or (ii) the 
 remaining data. The most accurate measurements are:
 ALEPH($\Afbl$), ALEPH(~\tpol), SLD($A_{LR}$), SLD jet charge
 ($A_{\rm{b}}$) and OPAL lepton ($\Afbb$). Although the weighted average error
  of the average using only the `most accurate' measurements is 70$\%$
larger than for all data, the resulting value of $A_{\rm{b}}=0.868(27)$ still 
 shows a -2.5$\sigma$ deviation from the SM. On the other hand, the
 remaining data with a weighted error only 38$\%$ larger than that
 for all data, gives a deviation of only -0.82$\sigma$ from the SM
prediction. The poor consistency between these two sets of data 
 evidently raises the question whether the systematic errors of
 some, or all, of 
 the `most accurate' measurements may have been under-estimated.
 If this is the case, the significance of the apparent deviation
 from the SM prediction may be much reduced.
\SECTION{\bf{The $A_{\ell}$ and $A_{\rm{b}}$ Measurements of the Different
 LEP and SLD Experiments}}
\begin{table}
\begin{center}
\begin{tabular}{|c||c|c|c|c|c|c|} \hline
  & $A_{\ell}$ & $A_{FB}^{0,\rm{b}}$ & $A_{\rm{b}}$ (WA $A_{\ell}$) & Dev($\sigma$) SM 
  &$A_{\rm{b}}$ (own $A_{\ell}$) & Dev($\sigma$) SM \\ \hline \hline
 ALEPH  & 0.1483(38) & 0.1040(35) & 0.931(33) & -0.12 & 0.935(40) &
0.0  \\ \hline
 DELPHI  & 0.1475(56) & 0.1006(41) & 0.900(38) & -0.92 & 0.909(50) &
-0.52 \\ \hline
 L3  & 0.1566(65) & 0.0956(62) & 0.855(56) & -1.4 & 0.814(63) &
-1.9  \\ \hline
OPAL  & 0.1350(62) & 0.0970(38) & 0.868(35) & -1.9 & 0.958(58) &
0.40  \\ \hline
SLD  & 0.1504(23) & - & - & - &0.908(27) &
-1.0  \\ \hline \hline
WA values & 0.1490(17) & 0.1002(21) & 0.896(19) & -2.1 & 0.911(18) &
-1.3 \\ \hline
\end{tabular}
\caption[]{ $A_{\ell}$ and $A_{\rm{b}}$ results of individual
 experiments. The last row shows Weighted Average (WA) values 
 calculated neglecting error correlations. The `own $A_{\ell}$' value
 for SLD refers
  to the direct measurement of $A_{\rm{b}}$ using the
 F/B-L/R asymmetry.}
    \end{center}
\end{table}
 The values of $A_{\ell}$ and $A_{\rm{b}}$ as measured separately by the
 four LEP experiments, and by SLD are presented in Table 5. For 
 each LEP experiment $A_{\rm{b}}$ is calculated in two different ways: (i)
 by use of the world average value of $A_{\ell}$ in Eqn.(1.2), or (ii)
 by use, instead, of the value of $A_{\ell}$  measured by the
 experiment itself. In each case the deviation of $A_{\rm{b}}$ from the
 SM prediction is shown. It may be noticed that, although ALEPH
 provides two out of the five `most accurate' measurements, that
 together 
 yield a -2.5$\sigma$ deviation from the SM (see Table 4), the
 ALEPH measurement itself, for both cases (i) and (ii), is in
 good agreement with the SM. DELPHI shows small deviations of
 -0.92$\sigma$, -0.52$\sigma$ in the cases (i) and (ii), whereas 
 L3 shows a larger deviation for case (ii) (-1.9$\sigma$) than 
 for case (i) (-1.4$\sigma$). An interesting case is OPAL, which 
 shows the largest deviation of any experiment (-1.9$\sigma$) in
 case (i), but a value quite consistent with the SM
 (0.40$\sigma$ deviation) in case (ii). This is easy to understand from
 Figs 1 and 2. The OPAL lepton measurement gives, as mentioned
 above, the most significant deviation of $A_{\rm{b}}$ from the SM
 for the case
 (i) (see Fig 1). However, it can seen in Fig 2 that the OPAL 
  values of $A_{\ell}$, as determined from $A_{FB}^{0,\ell}$ and the
 \tpol~measurement lie well below the WA value. The combined
 effect is so large, that for the case (ii), the deviations
 of $A_{FB}^{0,\rm{b}}$ and $A_{\ell}$ cancel almost exactly, giving an
 $A_{\rm{b}}$ value, calculated via Eqn(1.2), in agreement with the
 SM prediction.
\SECTION{\bf{The LEP and SLD Measurements of $A_{\rm{b}}$}}
 The separate LEP and SLD measurements of $A_{\rm{b}}$ are given in Table 2.
 They differ, respectively, from the SM prediction of 0.935 by 
 -2.3$\sigma$ and -1.0$\sigma$. The data are compared in more detail
 in Figs. 5, 6, 7 which show plots of the measured values of $A_{\rm{b}}$ 
 and $A_{\ell}$ for LEP, SLD and LEP+SLD respectively. In Figs 5 and 7 the
 LEP average $\Afbb$ measurement is shown as a diagonal band.
 In each case results of fits to $A_{\rm{b}}$ and $A_{\ell}$ are shown, as well
 as the SM prediction for a range of values of $m_t$ and $m_H$.
 In Figs 5 and 6 the dark square marked `WA' shows the World Average
 best fit value: $A_{\rm{b}} = 0.894$, $A_{\ell} = 0.1487$.
\begin{figure}[htbp]
\begin{center}\hspace*{-0.5cm}\mbox{
\epsfysize12.0cm\epsffile{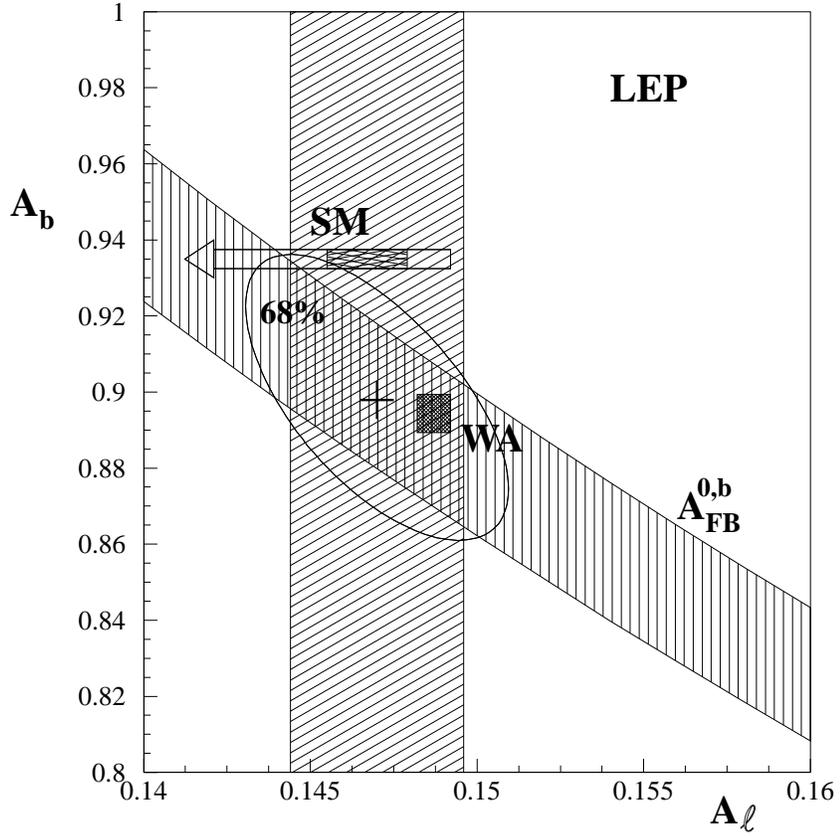}}
\caption{$A_{\rm{b}}$ versus $A_{\ell}$ plot for LEP data.The cross shows the
best fit value $A_{\ell}=0.1470$, $A_{\rm{b}}=0.898$, while the solid square 
marked WA (World Average) shows the result of the fit to the 
combined LEP+SLD data. The Standard Model prediction is
given by the arrow. The length of the shaft
(moving towards the tip) corresponds to a variation of
$m_H$ from 50 to 300 GeV ($m_t = 174$GeV) whereas the shaded
area corresponds to a variation of $m_t$ from 169 to 179 GeV
($m_H = 100$GeV). The 68$\%$ CL contour of the fit is also
 shown. }
\label{fig-fig5}
\end{center}
\end{figure}
\pagebreak 
\begin{figure}[htbp]
\begin{center}\hspace*{-0.5cm}\mbox{
\epsfysize12.0cm\epsffile{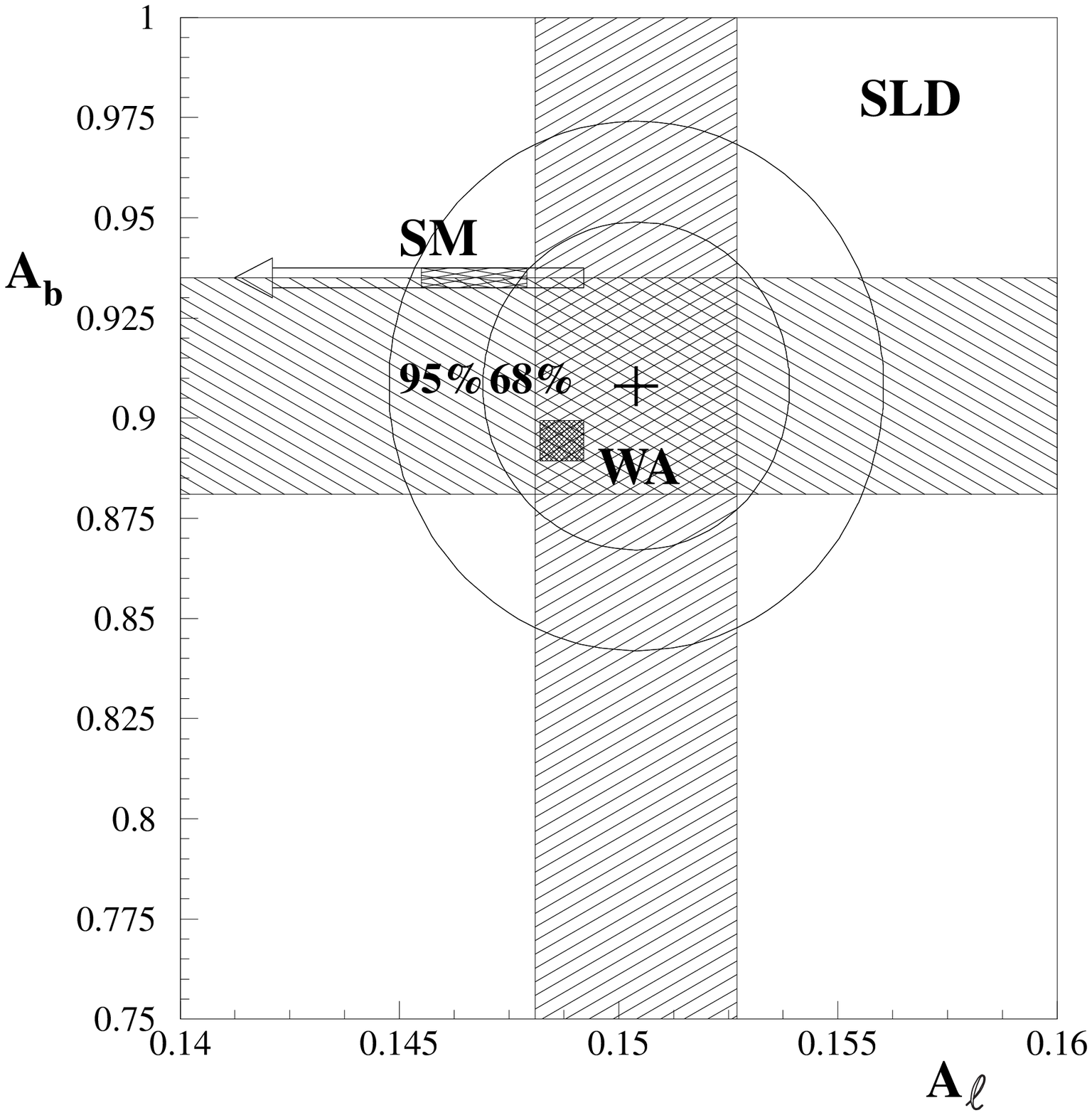}}
\caption{$A_{\rm{b}}$ versus $A_{\ell}$ plot for SLD data. The cross shows the
best fit value $A_{\ell}=0.1504$, $A_{\rm{b}}=0.908$, WA and the SM arrow 
are defined as in Fig 5. The 68$\%$ and 95$\%$ CL contours of
 the fit are shown.}
\label{fig-fig6}
\end{center}
\end{figure}
\pagebreak  
\begin{figure}[htbp]
\begin{center}\hspace*{-0.5cm}\mbox{
\epsfysize12.0cm\epsffile{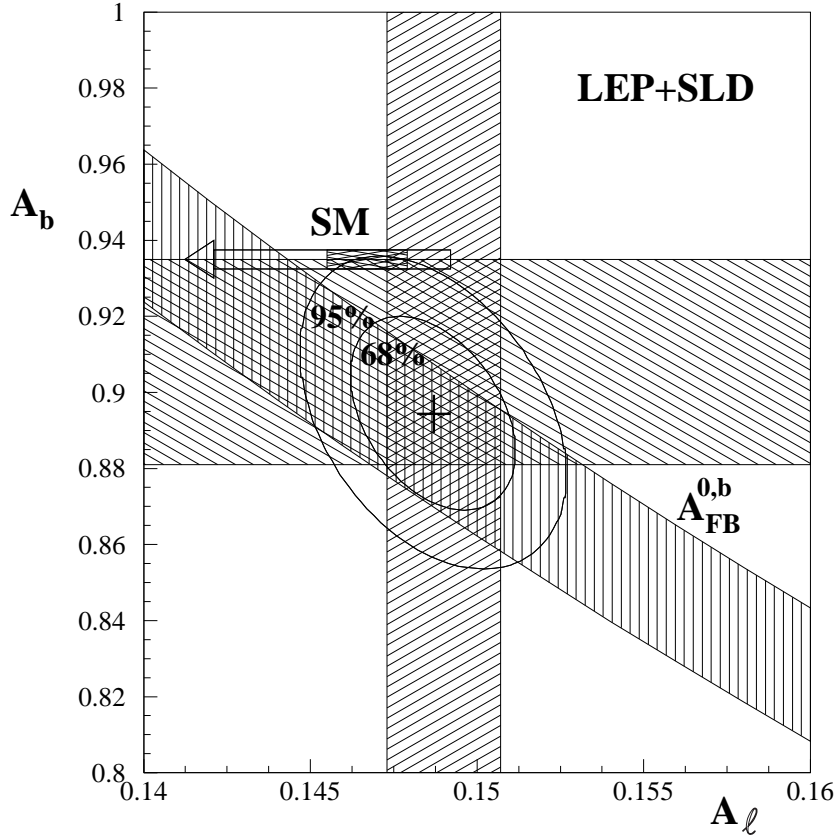}}
\caption{$A_{\rm{b}}$ versus $A_{\ell}$ plot for LEP+SLD data. The cross shows
 the best fit value $A_{\ell}=0.1487$, $A_{\rm{b}}=0.894$. WA and the SM arrow 
are defined as in Fig 5. The 68$\%$ and 95$\%$ CL contours of
 the fit are shown.}
\label{fig-fig7}
\end{center}
\end{figure} 
\SECTION{\bf{The Effect of Systematic
 Errors on the $A_{\rm{b}}$ Measurement}}
The different errors on the combined SLD and LEP measurements of
$A_{\rm{b}}$ as estimated by the LEP/SLD Heavy Flavour Working Group
are presented in Table 5~\cite{x14}. It can be seen that, even with
the full LEP1 data set of all four experiments, the error on 
the LEP average 
value remains statistics dominated, and that the systematic error is about
50$\%$ correlated. In contrast, the SLD statistical and
systematic errors are roughly equal and the correlated component of the
systematic error is relatively small. Since the forward/backward b quark
asymmetry measurements at SLD and LEP are very similar, and the 
 systematic error related to the beam polarisation measurement gives
 only a small contribution, it is reasonable to hope for a considerable
reduction in the SLD systematic error. Indeed, the smaller systematic
error at LEP is largely due to the much larger statistics of Z-decays
 at LEP, permitting systematic effects related to quark fragmentation
 to be estimated from the data itself.
\par Because of the large statistical weight of the LEP measurement,
 whose error is statistics dominated, the treatment of systematic
 errors is not expected to play a major r\^{o}le concerning the size
 of the $A_{\rm{b}}$ deviation. Even so, it is interesting to investigate 
 the effect of different treatments of the systematic error on the
 $A_{\rm{b}}$ deviation. It must not be forgotten that the estimation of 
 systematic errors is, perhaps, more of an art than a science, so
 that all confidence levels estimated on the assumption that the
 systematic errors are both correct and gaussian, should be taken
 {\it cum grano salis}. Here the effects are investigated
 of (i) using a uniform rather than a gaussian distribution
 for the systematic errors, (ii) an improvement in the systematic
 error of the SLD $A_{\rm{b}}$ measurement, (iii) optimism or 
 conservatism in the assignement of systematic errors.
\begin{table}
\begin{center}
\begin{tabular}{|c||c|c|} \hline
  & SLD & LEP \\ \hline \hline
$\sigma_{stat}$ & 0.017 & 0.019 \\ \hline
$\sigma_{syst}^{uncorr}$ & 0.019 & 0.007 \\ \hline
$\sigma_{syst}^{corr}$ & 0.0019 & 0.0061 \\ \hline
$\sigma_{tot}$ & 0.0349 & 0.0211 \\ \hline
\end{tabular}
\caption[]{ Statistical and systematic errors of the 
 combined SLD and LEP measurements of $A_{\rm{b}}$.}
\end{center}
\end{table}
\begin{table}
\begin{center}
\begin{tabular}{|c||c|} \hline
Syst. Error Hypothesis & $f$ $\%$ 
 \\ \hline \hline
Gaussian & 1.211(5) \\ \hline
Uniform & 1.188(5) \\ \hline
 LEP$\times$1.5 & 1.93(4) \\ \hline
 LEP/1.5 & 0.98(3) \\ \hline
 SLD/2.7 & 0.65(3) \\ \hline
 (SLD/2.7,LEP)$\times$1.5& 1.35(4) \\ \hline
(SLD/2.7,LEP)/1.5 & 0.44(5) \\  \hline
\end{tabular}
\caption[]{The effect of different hypotheses for
 systematic errors on the significance of the 
 observed $A_{\rm{b}}$ deviation. $f$ is the fraction of Monte Carlo ensembles of
 measurements with a simple average value of $A_{\rm{b}}$ less than the actual
 measured value (0.886).}
\end{center}
\end{table}  
\par A simple Monte Carlo program was used to generate
ensembles of $A_{\rm{b}}$ measurements distributed 
 according to the statistical and systematic 
errors of the different LEP and SLD experiments as shown in Tables 2 
and 6. The correlated and uncorrelated components of the different
$A_{\rm{b}}$ and $A_{FB}^{0,\rm{b}}$ measurements were
properly taken into account. In all cases
except one (see below) the systematic errors were modeled according to 
gaussian functions with RMS equal to the quoted errors. The error on the
LEP+SLD average value of $A_{\ell}$ used to extract the LEP values of $A_{\rm{b}}$
according to Eqn(1.2) was taken to be gaussian and 100$\%$ correlated
between the different measurements. The Standard Model value of
$A_{\rm{b}}$ (0.935) was assumed, and the fraction, $f$, of ensembles of
measurements with
a simple mean value of $A_{\rm{b}}$ less than that given by the data
($A_{\rm{b}}=0.886$) was noted. In Table 6 the values of $f$ (corresponding to
a one-sided CL) are shown for several different hypotheses concerning 
the errors. The first row corresponds to the quoted errors and assumes
gaussian distributions. In the second row, all systematic errors are
chosen according to uniform distributions with RMS equal to the
quoted errors. In the third (fourth) rows the effect is shown of
increasing (decreasing) the systematic errors of all the LEP experiments
by a factor 1.5. In the fifth row is shown the effect of reducing the
systematic errors of the SLD experiments by a factor 2.7 so that 
the average systematic error becomes equal to the 
uncorrelated  LEP systematic error. Finally, in
the last two rows an additional scale factor of 1.5 or 1/1.5 is applied
to the systematic errors of all experiments. As anticipated above,
 different scenarios for the systematic errors do not have a dramatic 
 effect on the significance of the observed deviation. Use of a uniform
 distribution instead of a gaussian one (expected to reduce the tails of
 distribution) in fact only gives a 2$\%$ relative change in $f$.
 Assuming that the SLD systematic error is reduced to the same level
 as the current LEP one, overestimation (underestimation) of all
 systmatic errors by a factor 1.5 gives CLs of 0.44$\%$ (1.4$\%$) that
 the observed fluctuation is purely statistical, to be compared
 with 1.2$\%$ for the nominal errors. 
\par It may finally be remarked that a previous study~\cite{x4} of
 Z decay measurements showed a clear tendency to overestimate 
 point-to-point systematic errors and to underestimate correlated
 ones. Correcting for the first effect would increase the significance
 of any deviation,
 while correcting for the second would tend to decrease it.
 Unfortunately, there are 
 insufficient independent measurements to perform a similar 
 analysis in the present case.

\SECTION{\bf{Summary and Outlook}}
 This paper has studied, in detail, the data dependence of the
 parameter $A_{\rm{b}}$. The individual measurements of both $A_{\rm{b}}$ and
 the related (for LEP) parameter $A_{\ell}$ show quite good internal 
 consistency. For $A_{\rm{b}}$ the largest positive deviation from the
 WA is given by the ALEPH jet charge measurement. Removing this
 increases the $A_{\rm{b}}$ deviation from -2.4$\sigma$ to -2.8$\sigma$.
 The single measurement with the largest weight tending to 
 increase the size of the deviation is the OPAL lepton
 $\Afbb$ measurement. Removing this reduces the $A_{\rm{b}}$
 deviation to -1.7$\sigma$. For $A_{\ell}$ it may be noted that the
 $A_{FB}^{0,\ell}$ measurement of OPAL and the \tpol~ measurements
 of DELPHI and OPAL all lie about 2$\sigma$ below the WA. Excluding
 these measurements slightly increases the $A_{\rm{b}}$ deviation to
-2.6$\sigma$ .
 The deviation observed is much larger (-2.5$\sigma$) if only the
 most accurate measurements of each raw observable are used, than for
 all the remaining measurements (-0.82$\sigma$). This is a possible
 hint that the systematic errors of the `most accurate' measurements
 may be underestimated, leading to an overestimation of the deviation
 from the SM for these data. The independent measurements of $A_{\rm{b}}$ for
 each LEP experiment give smaller deviations for all experiments, except
 L3, than when the world average value of $A_{\ell}$ is used to extract $A_{\rm{b}}$.
 The naive WA (neglecting error correlations) of the individual measurements
 of $A_{\rm{b}}$ of the four LEP experiments and SLD shows only a -1.3$\sigma$
 deviation. Using the
 world average value of $A_{\ell}$ to extract $A_{\rm{b}}$ from the LEP
 experiments yields a deviation of -2.1$\sigma$ to be compared with
 -1.0$\sigma$ for the combined SLD experiments. A study of the
 modelling and the degree of optimism/conservatism in the estimation 
 of systematic errors shows essentially identical results for 
 gaussian or uniform distributions and values for the CL for agreement
 with the SM that
  varies from 0.44$\%$  to 1.9$\%$, as compared to the
 nominal value of 1.2$\%$.
\par In the future, some improvement may be expected in the SLD
 values of $A_{\ell}$ and $A_{\rm{b}}$, mainly due to an improved understanding
 of systematic errors~\cite{x15}. On the other hand, no significant
 improvement is to be expected from the LEP results which, although
 many are still `preliminary', are almost
 entirely based on the full LEP1 statistics. It may be noted that 
 a recent summary of the SLD data~\cite{x15}
 found slightly different values
 for the LEP,SLD average values of $A_{\rm{b}}$ of 0.877(21), 0.898(29)
 respectively (compare with the values given in Table 2). The 
 small differences from the values used above do not affect any
 of the conclusions of this study.
\par This paper is based on the precision electroweak data 
available in Spring 1999. In the Summer 1999 update~\cite{x16}, the
 values
 0.881(20), 0.905(26) were given for the LEP, SLD average values, 
 respectively, of $A_{\rm{b}}$. A fit to the combined LEP+SLD data for
$A_{\ell}$ and $A_{\rm{b}}$, similar to those shown if Figs.(5-7) of this paper,
yielded the values; $A_{\ell} = 0.1493(16)$, $A_{\rm{b}} = 0.889(16)$. Thus,
 in the most recent data, the significance of the $A_{\rm{b}}$ deviation has
 increased to 2.9$\sigma$. 
 \par Finally, the deviation in $A_{\rm{b}}$ although interesting, and 
 possibly suggestive of new physics~\cite{x17,x18,x19} is still of
 only marginal statistical significance. If there is no fresh data
 from SLD it may be some decades before it is known for sure
 if the
 effective couplings of the b quarks are, or are not, in agreement
 with the SM predictions!
\newline
\newline
{\bf Acknowledgements}
\par We thank Simon Blyth and Michael Dittmar for their careful readings
 of the paper and their helpful comments, and Franz Muheim for 
 discussions of the LEP/SLD Heavy Flavour Working Group averages.      
\pagebreak


\begin{thebibliography}{99}
\bibitem{x1}
The LEP Collaborations ALEPH, DELPHI, L3, OPAL,
the LEP Electroweak Working Group and the SLD
Heavy Flavour Group.`A Combination of Preliminary Measurements
 and Constraints on the Standard Model' (EWWGR),
 CERN-PPE/96-183 (1996). 
\bibitem{x2}
P.B.Renton, Int. Journ. Mod. Phys. {\bf A12}, 4109 (1997);
\newline
 Eur. Phys. J {\bf C8} (1999).
\bibitem{x3}
J.H.Field, Mod. Phys. Lett. A, Vol. 13 No. 24, 1937 (1998);
\newline
Vol. 14, No. 26, 1815 (1999),
\bibitem{x4}
J.H.Field, Phys. Rev. {\bf D58} 093010-1 (1998).
\bibitem{x5}
S.L.Glashow, Nucl. Phys. {\bf 22}, 579 (1961). 
S.Weinberg Phys. Rev. Lett. {\bf 19}, 1264 (1967). 
A.Salam in `Elementary Particle Theory' Ed. N.Svartholm,
Stockholm 1968, P367.
\bibitem{x6}
G.Rodrigo, Nucl. Phys. B (Proc. Suppl.) {\bf 54A}, 60 (1997).
\bibitem{x7}
 EWWGR,CERN-PPE/93-157 (1993).
\bibitem{x8}
 EWWGR,CERN-PPE/94-187 (1994).
\bibitem{x9}
 EWWGR,CERN-PPE/95-172 (1995).
\bibitem{x10}
 EWWGR,CERN-PPE/97-154 (1997).
\bibitem{x11}
M.Gr\"{u}newald, in proceedings of the XXIX International
 Conference on High Energy Physics, UBC, Vancouver, BC, Canada, 
July 23-29 1998.   
\bibitem{x12}
Sal Fahey, in proceedings of the XXIX International
 Conference on High Energy Physics, UBC, Vancouver, BC, Canada, 
July 23-29 1998. 
\bibitem{x13}
Martin  Gr\"{u}newald and Klaus M\"{o}nig, private communications.
\bibitem{x14}
Franz Muheim, private communication.
\bibitem{x15}
 P.C.Rowson, SLAC-PUB-8132, April 1999.
 e-print hep-ex/9904016.
\bibitem{x16}
M. Swartz, in proceedings of the XIX International
Symposium on Lepton and Photon Interactions at High Energies,
Stanford University, California, U.S.A., August 9-14 1999. 
\bibitem{x17}
D.Chang, W-F Chang and E.Ma, Phys. Rev. {\bf D59} 091503 (1999);
UCRHEP-T263, e-print hep-ph/9909537.
\bibitem{x18}
K.Y.Lee, SNUTP 99-017, e-print hep-ph/9904435.
\bibitem{x19}
M.S.Chanowitz, LBNL-43248, e-print hep-ph/9905478. 
\end{thebibliography}
\end{document}